\documentclass[12pt,a4paper,onecolumn,draftcls]{IEEEtran}
\usepackage[latin1]{inputenc}
\usepackage{graphicx}
\usepackage{amssymb,amsmath,amsfonts,amsthm}
\usepackage{algorithm,algorithmic}
\usepackage{cite}
\usepackage{float}
\usepackage{graphics}
\usepackage{subfigure}
\usepackage{xspace}
\usepackage[usenames,dvipsnames]{color}
\usepackage{amssymb, epsfig, hyperref}
\usepackage{mathtools}
\usepackage{bbm}
\usepackage{subeqnarray}

\begin{document}
\title{Simultaneous Wireless Information and Power Transfer in Modern \\ Communication Systems}

\author{Ioannis Krikidis,~\IEEEmembership{Senior Member},~IEEE, Stelios Timotheou,~\IEEEmembership{Member},~IEEE, Symeon Nikolaou,~\IEEEmembership{Member},~IEEE, Gan Zheng,~\IEEEmembership{Senior Member},~IEEE, Derrick Wing Kwan Ng,~\IEEEmembership{Member},~IEEE, and Robert Schober,~\IEEEmembership{Fellow},~IEEE
\thanks{I. Krikidis is with the Department of Electrical and Computer Engineering,
University of Cyprus, Cyprus (E-mail: krikidis@ucy.ac.cy).}
\thanks{S. Timotheou is with the KIOS Research Center, University of Cyprus, Cyprus (E-mail: timotheou.stelios@ucy.ac.cy).}
\thanks{S. Nikolaou is with the Department of Electrical Engineering, Frederick University, Cyprus (E-mail: eng.ns@frederick.ac.cy).}
\thanks{G. Zheng is with the School of Computer Science and Electronic Engineering, University of Essex, UK (E-mail: ganzheng@essex.ac.uk).}
\thanks{D. W. K. Ng and R. Schober are with the Institute for Digital Communications
(IDC), Friedrich-Alexander-University Erlangen-Nurnberg (FAU),
Germany (E-mail: \{kwan, schober\}@lnt.de).}}

\maketitle

\section{Abstract}
Energy harvesting for wireless communication networks is a new paradigm that allows terminals to recharge their batteries from external energy sources in the surrounding environment. A promising energy harvesting technology is wireless power transfer where terminals harvest energy from electromagnetic radiation. Thereby, the energy may be harvested opportunistically from ambient electromagnetic sources or from sources that intentionally transmit electromagnetic energy for energy harvesting purposes. A particularly interesting and challenging scenario arises when sources perform simultaneous wireless information and power transfer (SWIPT), as strong signals not only increase power transfer but also interference. This paper provides an overview of SWIPT systems with a particular focus on the hardware realization of rectenna circuits and practical techniques that achieve SWIPT in the domains of time, power, antennas, and space. The paper also discusses the benefits of a potential integration of SWIPT technologies in modern communication networks in the context of resource allocation and cooperative cognitive radio networks.

\section{Introduction}

Recently, there has been a lot of interest to integrate energy harvesting technologies into communication networks. Several studies have considered conventional renewable energy resources, such as solar, wind etc, and have investigated optimal resource allocation techniques for different objective functions and topologies. However, the intermittent and unpredictable nature of these energy sources makes energy harvesting critical for applications where quality-of-service (QoS) is of paramount importance, and most conventional harvesting technologies are only applicable in certain environments. An energy harvesting technology that overcomes the above limitations, is wireless power transfer (WPT), where the nodes charge their batteries from electromagnetic radiation. In WPT, green energy can be harvested either from ambient signals opportunistically, or from a dedicated source in a fully-controlled manner; in the latter case, green energy transfer can take place from more powerful nodes  (e.g. base stations) that exploit conventional forms of renewable energy.

Initial efforts on WPT have focused on long-distance and high-power applications. However, both the low efficiency of the transmission process and health concerns for such high-power applications prevented their further development. Therefore, most recent WPT research has focused on near-field energy transmission through inductive coupling  (e.g., used for charging cell-phones, medical implants, and electrical vehicles). In addition, recent advances in silicon technology have significantly reduced the energy demand of simple wireless devices. WPT is an innovative technology and attracts the interest from both the academia and the industry; some commercial WPT products already exist e.g. \cite{PCAST} and several experimental results for different WPT scenarios are reported in the literature \cite{SHI}. With sensors and wireless transceivers getting ever smaller and more energy efficient, we envision that radio waves will not only become a major source of energy for operating these devices, but their information and energy transmission aspects will also be unified. Simultaneous wireless information and power transfer (SWIPT) can result in significant gains in terms of spectral efficiency, time delay, energy consumption, and interference management by superposing information and power transfer. For example, wireless implants can be charged and calibrated concurrently with the same signal and wireless sensor nodes can be charged with the control signals they receive from the access point. In the era of Internet of Things, SWIPT technologies can be of fundamental importance for energy supply to and information exchange with numerous ultra-low power sensors, that support heterogeneous sensing applications. Also, future cellular systems with small cells, massive multiple-input multiple-output (MIMO) and millimeter-wave technologies will overcome current path-loss effects; in this case,  SWIPT could be integrated as an efficient way to jointly support high throughputs and energy sustainability.

In this paper, we give an overview of the SWIPT technology and discuss recent advances and future research challenges. More specifically, we explain the rectenna (rectifying antenna) circuit which converts microwave energy into direct current (DC) electricity and is an essential block for the implementation of the WPT/SWIPT technology.  Due to practical limitations, SWIPT requires the splitting of the received signal in two orthogonal parts. Recent SWIPT techniques that separate the received signal in the domains of time, power, antenna, and space are presented. On the other hand, SWIPT entails fundamental modifications for the operation of a communication system and motivates new applications and services. From this perspective, we discuss the impact of SWIPT on the radio resource allocation problem as well as sophisticated cognitive radio (CR) scenarios which enable information and energy cooperation between primary and secondary networks.

\section{WPT module components}

Exchanging electromagnetic power wirelessly can be classified into three distinct cases: a) Near field power transfer employing inductive, capacitive or resonant coupling that can transfer power in the range of tenths of Watts, over short distances of up to one meter (sub-wavelength). b) Far field directive power beaming, requiring directive antennas, that can transfer power in the range of several mWatts at distances of up to several meters in indoor and outdoor environments. c) Far field, low-power, ambient RF power scavenging  involving receivers that opportunistically scavenge the power transmitted from public random transmitters (cell phone base stations, TV broadcasting stations) for their communication with their peer nodes. For this last case the collected power is in the range of several $\mu$Watts, and the communication range can be up to several km assuming there is adequate power density.  While there are several applications related to near field wireless charging, such as wireless charging of electric cars, cell phones or other hand-held devices, the main focus of this paper will be on far field WPT which involves the use of antennas communicating in the far field.

\subsection{Wireless Power Receiver Module}
\begin{figure}[t]
\centering
\includegraphics[width=0.6\linewidth]{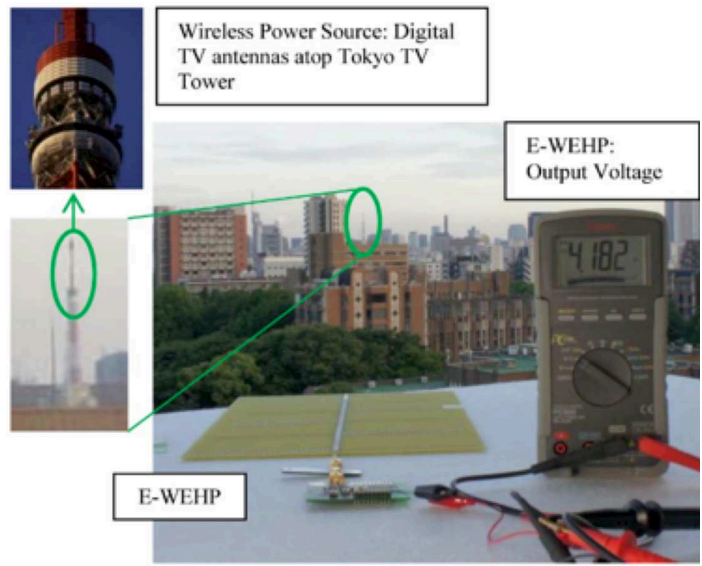}
 \caption{Field measurement in downtown Tokyo, Japan, with prototype device \cite{TEB} harvesting wireless energy from multi-carrier wireless digital TV signals broadcasted from atop the Tokyo TV tower $6.3$ km away.} \label{rectenna2}\vspace*{-4mm}
\end{figure}

A wireless power scavenger or receiver consists of the following components: A receiver antenna or antenna array, a matching network, a radio frequency to direct current (RF-DC) converter or rectifier, a power management unit (PMU) and the energy storage unit \cite{POP}. Upon the successful charging of the energy storage unit, the storage unit, usually a rechargeable battery or a super capacitor, will provide power to the central processing unit (CPU), the sensors and the low duty cycle communication transceiver.  The schematic of this module is presented in Fig. \ref{rectenna1} and a successful implementation of a WPT system that scavenges ambient power $6.3$ km away from Tokyo TV tower is shown in Fig. \ref{rectenna2}.

\subsection{Conditions for Efficient WPT}
Based on Friis free space equation the received RF power at the terminals of the antenna depends on the available power density and the antennas' effective area $A_e=(\lambda^2 G_R)/(4\pi)$ and is given by:

\begin{figure}[t]
\centering
\includegraphics[width=0.6\linewidth]{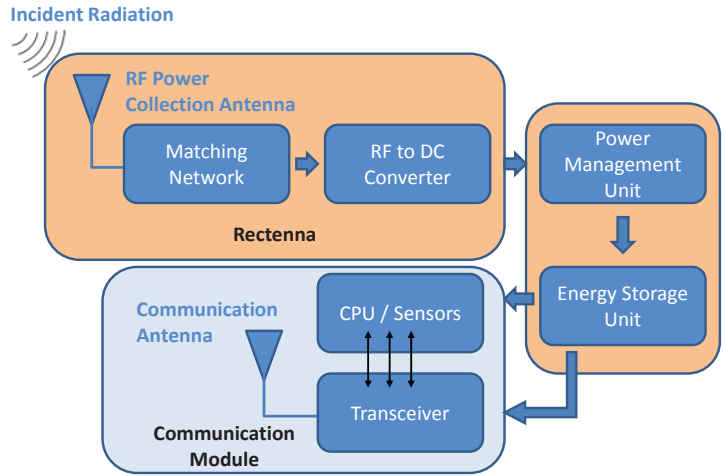}
 \caption{Block diagram of a typical power scavenging module powering a communication transceiver.} \label{rectenna1}\vspace*{-4mm}
\end{figure}

\begin{align}\label{friis}
P_R=\cos^2\phi \frac{P_T G_T}{4\pi R^2}A_e,
\end{align}
where $P_T$ and $P_R$ are the transmitted and received power, respectively, $G_T$ and $G_R$ are the transmitter and receiver gains (functions of the spatial variables) respectively, $\lambda$ denotes the wavelength, and $\cos \phi$ is the polarization loss factor which accounts for the misalignment (angle $\phi$) of the received  electric intensity vector $E$ and the receiver antenna linear polarization vector.  From \eqref{friis} we can deduce that in order to ensure maximum received power, the receiver antenna needs to have high gain, it has to be directed towards the transmitter (maximum directivity direction), and it has to be aligned with the received $E$-field ($\phi=0$). However, these conditions cannot be ensured in practice. For example, in a Rayleigh multipath propagation environment the received signal has random polarization. Consequently the optimum polarization for a receiver antenna is dual, linear, orthogonal polarization because it ensures the reception of the maximum average power regardless the received signal's polarization. If the maximum gain direction cannot be guaranteed, omni-directional antennas are preferred instead. Friis equation is frequency dependent and is applicable to narrowband signals. The total received power is calculated by integrating the received power $P_R$ over frequency, therefore, a broadband antenna will receive more power than a narrowband one. As a result, wideband antennas or multi-band antennas are preferred.

The RF-to-DC converter or rectifier is probably the most critical component of a WPT module and its design is the most challenging task \cite{NIN}. A rectifier consists of at least one non-linear device. Most rectennas (antenna and rectifier co-design) reported in literature consist of only one diode. Ideally, the conversion efficiency of a rectifying circuit with a single non-linear device, can reach up to $100$\%. Unfortunately, this can only happen for specific values of $P_{RF}$ and $R_{DC}$, where $P_{RF}$ denotes the level of the input RF power at the rectifier, and $R_{DC}$ is the delivered load. In more detail, the rectenna structure consists of a single shunt full-wave rectifying circuit with one diode, a $\lambda/4$ distributed line, and a capacitor to reduce the loss in the diode. Depending on the requirements, more complicated and sophisticated rectifier topologies can be used which are based on the well known Dickson charge pump that can provide both rectification and impedance transformation. Typically, Schottky diodes are used as the non-linear devices  because they have low forward voltage drop and allow very fast switching action, features useful for rectifiers. Low forward voltage drop is needed because the received power is rather small, and fast switching action is needed to follow the relatively high RF frequency of the received signal. Alternatively, it is possible to use CMOS transistors or other transistors as the non-linear rectifying elements especially when integrated solutions are preferred. The major problem with RF-to-DC converters is that their efficiency, defined as $n_R=P_{RF}/(V_{DC}^2/R_{DC})$  depends on $P_{RF}$, $R_{DC}$, and the DC voltage, $V_{DC}$, across the load. Generally the higher the incident RF power the higher the efficiency. For low power levels, efficiency can even drop to zero because the diodes'  forward voltage drop is too high. This is why the reported high efficiencies cannot be seen in actual RF scavenging scenarios. As an example, the ambient power density measured $6.5$ km far from the Tokyo TV tower was approximately $1$ $\mu$W/cm$^2$ and the received power was about $50$ $\mu$W whereas high efficiency rectifiers require input powers between $0.5-5$ mW, ten to a hundred times higher. As a result, the measured efficiency was rather small.

The final stage of the WPT module is the Power Management Unit (PMU) that is responsible for maintaining the optimum load at the terminals of the rectifier despite the changing received RF power levels, and at the same time ensures the charging of the Energy Storage Unit without additional loss.

\section{Techniques for SWIPT}

Early information theoretical studies on SWIPT have assumed that the same signal can convey both energy and information without losses, revealing a fundamental trade-off between information and power transfer \cite{CN:Shannon_meets_tesla}.  However, this simultaneous transfer is not possible in practice, as the energy harvesting operation performed in the RF domain destroys the information content. To practically achieve SWIPT, the received signal has to be split in two distinct parts, one for energy harvesting and one for information decoding. In the following, the techniques that have been proposed to achieve this signal splitting in different domains (time, power, antenna, space) are discussed.

\begin{figure}[t]
\centering
\includegraphics[width=0.8\linewidth]{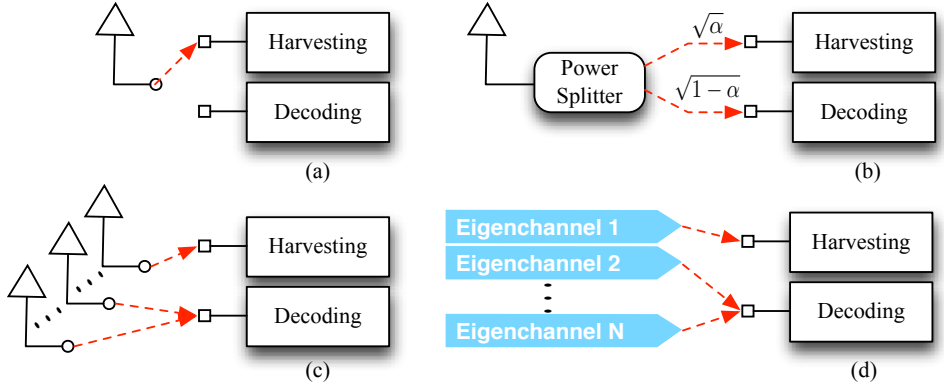}
\caption{SWIPT transmission techniques in different domains: a) time, b) power, c) antenna, and d) space; $\alpha$ denotes the PS factor.}\label{model1}
\end{figure}

\subsection{Time Switching (TS)}
If TS is employed, the receiver switches in time between information decoding and energy harvesting \cite{RUI1}. In this case, the signal splitting is performed in the time domain and thus the entire signal received in one time slot is used either for information decoding or power transfer (Fig. \ref{model1}a). The TS technique allows for a simple hardware implementation at the receiver but requires accurate time synchronization and information/energy  scheduling.
\subsection{Power Splitting (PS)}
The PS technique achieves SWIPT by splitting the received signal in two streams of different power levels using a PS component; one signal stream is sent to the rectenna circuit for energy harvesting and the other is converted to baseband for information decoding (Fig. \ref{model1}b) \cite{RUI1}. The PS technique entails a higher receiver complexity compared to TS and requires the optimization of the PS factor $\alpha$; however, it achieves instantaneous SWIPT, as the signal received in one time slot is used for both information decoding and power transfer. Therefore, it is more suitable for applications with critical information/energy or delay constraints and closer to the information theoretical optimum.

\subsection{Antenna Switching (AS)}
\label{sec:AS}
Typically, antenna arrays are used to generate DC power for reliable device operation. Inspired by this approach,  the AS technique dynamically switches each antenna element between decoding/rectifying to achieve SWIPT in the antenna domain (Fig. \ref{model1}c). In the AS scheme, the receiving antennas are divided into two groups where one group  is used for information decoding and the other group for energy harvesting \cite{RUI1}. The AS technique requires the solution of an optimization problem in each communication frame in order to decide the optimal assignment of the antenna elements for information decoding and energy harvesting. For a MIMO decode-and-forward (DF) relay channel, where the relay node uses the harvested energy in order to retransmit the received signal, the optimization problem was formulated as a knapsack problem and solved using dynamic programming in \cite{KRI}.

Because optimal AS suffers from high complexity, low-complexity AS mechanisms have been devised which use the principles of generalized selection combining (GSC) \cite{KRI}. The key idea of GSC-AS is to use the $L$ out of $N_T$ antennas with the strongest channel paths for either energy (GSCE technique) or information (GSCI technique) and the rest for the other operation.
\subsection{Spatial Switching (SS)}
The SS technique can be applied in MIMO configurations and achieves SWIPT in the spatial domain by exploiting the multiple degrees of freedom (DoF) of the interference channel \cite{TIM}.  Based on the singular value decomposition (SVD) of the MIMO channel, the communication link is transformed into parallel eigenchannels that can convey either information or energy (Fig. \ref{model1}d).  At the output of each eigenchannel there is a switch that drives the channel output either to the conventional decoding circuit or to the rectification circuit. Eigenchannel assignment and power allocation in different eigenchannels is a difficult nonlinear combinatorial optimization problem; in \cite{TIM} an optimal polynomial complexity algorithm has been proposed for the special case of unlimited maximum power per eigenchannel.

 {\it Numerical example:} The performance of the discussed SWIPT techniques is illustrated for the MIMO relay channel introduced in Section \ref{sec:AS} assuming a normalized block fading Rayleigh. In the considered set-up, a single-antenna source communicates with a single-antenna destination through a battery-free MIMO relay node, which uses the harvested energy in order to power the relaying transmission. We assume that the source transmits with power $P$ and spectral efficiency $r_0=2$ bits per channel use (BPCU); the relay node has global channel knowledge, which enables beamforming for the relaying link. An outage event occurs when the destination is not able to decode the transmitted signal and the performance metric is the outage probability. The first observation is that GSCI outperforms GSCE scheme for $L=1$ and $L=2$, respectively. This result shows that diversity gain becomes more important than energy harvesting due to the high RF-to-DC efficiency $\eta$. In addition,  the GSCI scheme with $L=1$ is the optimal GSC-based strategy and achieves a diversity gain equal to two. It can be also seen that the PS scheme outperforms the AS scheme with a gain of $2.5$ dB for high $P$, while the TS scheme provides a poor performance due to the required time division.

\begin{figure}[t]
\centering
\includegraphics[width=0.6\linewidth]{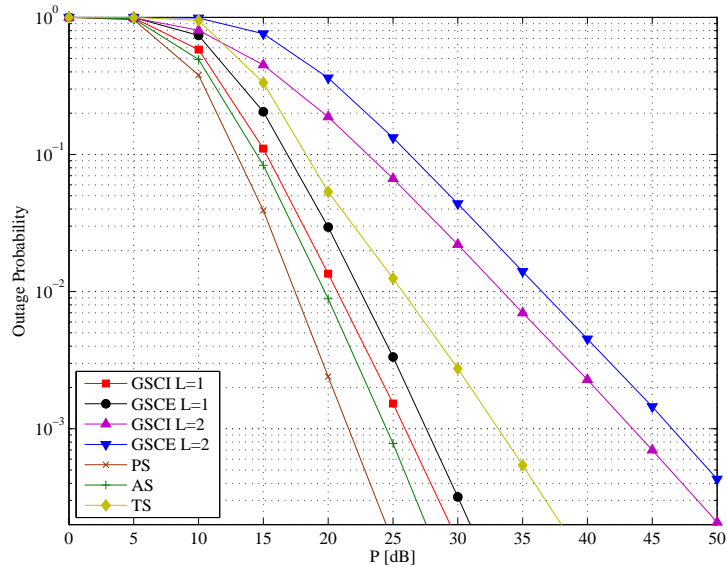}
\caption{Outage probability versus $P$ for GSCI, GSCE, PS, AS and TS; the simulation setup is $r_0=2$ BPCU, $N_T=3$ antennas, $L=\{1,2\}$ and RF-to-DC efficiency $\eta=1$.}\label{figure2}
\end{figure}

\section{Resource Allocation for Systems with SWIPT}
This section discusses the benefits of employing SWIPT on resource allocation applications. Utility-based resource allocation algorithm design has been heavily studied in the literature \cite{JR:resource_allocation} for optimizing the utilization of limited resources in the physical layer such as energy, bandwidth, time, and
space in multiuser systems.   In addition to the conventional  QoS requirements  such as throughput, reliability, energy efficiency, fairness, and delay, the efficient  transfer of energy plays an important role as a new QoS requirement for SWIPT \cite{CN:Shannon_meets_tesla,JR:WIPT_fullpaper}. Resource allocation algorithm design for SWIPT systems includes the following aspects:

\begin{figure}[t]
\centering
\includegraphics[width=0.6\linewidth]{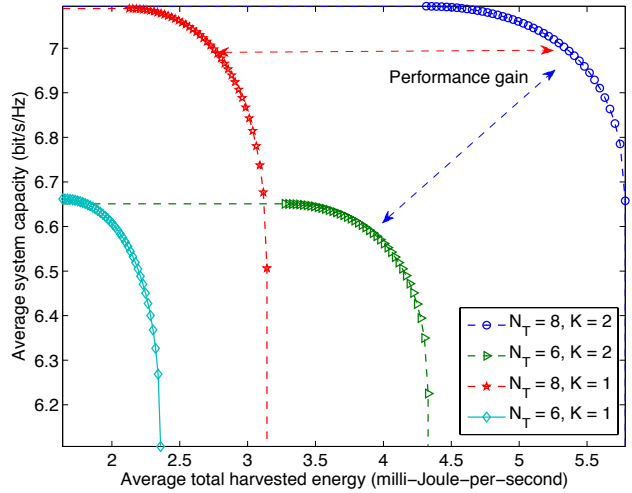}
 \caption{The trade-off region of the average system capacity (bit/s/Hz) and the average total harvested energy (milli-Joule-per-second)  for different numbers of receivers. The carrier frequency is $915$ MHz and the information receiver and energy harvesting receivers are located at $30$ meters and $10$ meters from the transmitter, respectively. The total transmit power, noise power, transceiver antenna gain, and RF-to-DC conversion loss are set to $10$ Watt, $-23$ dBm, $10$ dBi, and $3$ dB, respectively.} \label{fig:cap_EH}\vspace*{-4mm}
\end{figure}

\begin{itemize}
\item Joint power control and user scheduling -- The RF signal acts as a dual purpose carrier for conveying information and energy to the receivers simultaneously. However,  the wide dynamic range of the power sensitivity for energy harvesting ($-10$ dBm) and information decoding ($-60$ dBm) is an obstacle for realizing SWIPT. As a result,  joint power control and user scheduling is a key aspect for  facilitating  SWIPT in practice. For instance,  idle users experiencing  high  channel gains can be scheduled for power transfer to extend the life time of the communication network. Besides, opportunistic power control can be used to exploit the channel fading   for improved  energy and information transfer efficiency \cite{CN:Shannon_meets_tesla, JR:WIPT_fullpaper, JR:Kwan_secure_imperfect}. Fig. \ref{fig:cap_EH} depicts an example of power control in SWIPT systems.  We show the average system capacity versus the average total harvested energy in a downlink system. In particular,  a transmitter equipped with $N_{\mathrm{T}}$ antennas is serving one single-antenna information receiver and $K$ single-antenna energy harvesting receivers. As can be observed, with optimal power control, the trade-off region of the system capacity and the harvested energy increases significantly  with $N_{\mathrm{T}}$.  Besides, the average harvested energy improves with the number of energy harvesting receivers.

\item Energy and information scheduling  -- For passive receivers such as small sensor nodes, uplink data transmission is only  possible after the receivers have harvested a sufficient amount of energy from  the RF in the downlink. The physical constraint on the energy usage motivates a ``harvest-then-transmit" design. Allocating more time for energy harvesting in the downlink leads to a higher amount of harvested energy which can then be used in the uplink. Yet,  this also implies that there is less time for uplink transmission which may result in a lower  transmission data rate.   Thus, by varying the amounts of time allocated for energy harvesting and information transmission, the system throughput can be optimized.

\item Interference management -- In traditional communication networks, co-channel interference is recognized as one of the major factors that limits the system performance and is suppressed or avoided via resource allocation. However, in SWIPT systems, the receivers may embrace strong interference since it can act as a vital source of energy. In fact, injecting artificial interference into the communication network may be beneficial  for the overall system performance, especially when the receivers do not have enough energy for supporting their normal operations, since in this case, information decoding becomes less important compared to energy harvesting. Besides, by exploiting interference alignment and/or interference coordination, a ``wireless charging zone" can be created by concentrating and gathering  multicell interference in certain locations.
\end{itemize}

\section{Joint Information and Energy Cooperation in  CR Networks}

SWIPT also opens up new opportunities for cooperative communications. We present one example where  SWIPT improves the traditional system design
of cooperative CR networks (CCRNs). CCRNs are a new paradigm for
improving the spectrum sharing by having the primary and secondary
systems actively seek opportunities to cooperate with each other
\cite{Spectrum-leasing-Simeone}\cite{Zheng_CCRN_13}. The secondary
transmitter (ST) helps in relaying the traffic of the primary transmitter
(PT) to the primary user (PU), and in return can utilize the primary
spectrum to serve its own secondary user (SU). However, to enable
this cooperation, the ST should both possess a good channel link to
the primary system and have sufficient transmit power. While the
former can be achieved by proper placement, the latter requirement
cannot be easily met especially  when the ST is a low-power relay
node rather than  a powerful base station (BS), which renders this
cooperation not meaningful.

SWIPT could provide a promising solution to
address this challenge by encouraging  the cooperation between the
primary and  secondary systems at both the information and the
energy levels \cite{Zheng_2014_CR_EH}, i.e., the PT will transmit
both information and energy to the ST, in exchange, the low-power ST
relays the primary information.  Compared to the traditional CCRN,
this approach creates more incentives for
 both systems to cooperate and  therefore  improves the system overall spectrum
 efficiency without relying on external energy sources.

We illustrate the performance gain by studying a joint information
and energy cooperation scheme using the amplify-and-forward  protocol and the power splitting technique. Two channel phases are required to complete the communication. In Phase I, the PT broadcasts its data and both the ST and the PU listen. The ST then splits the received RF signal into two parts: one for information processing then forwarding to the PU and the other for harvesting energy, with relative power ratio of $\alpha$ and $1-\alpha$,
respectively. In Phase II, the ST superimposes the processed primary
data with its own precoded data, then transmits it to both the PU
and the SU. The ST jointly optimizes power allocation factor
$\alpha$ and the precoding vectors to the PU and SU to achieve the
maximum rates.

\begin{figure}[t]
\centering
\includegraphics[width=3.5in]{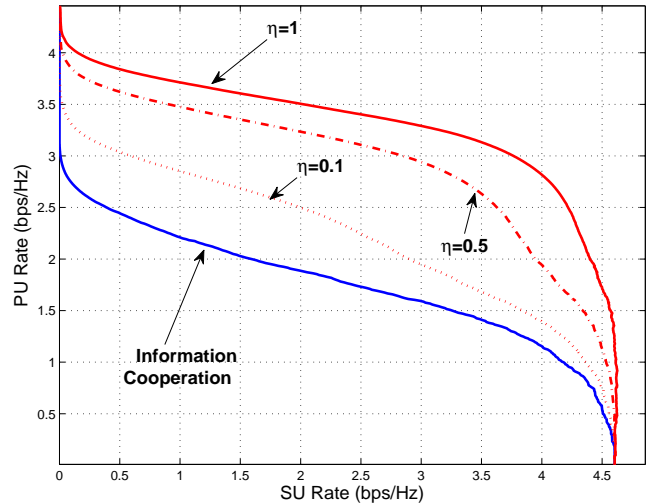}
\caption{PU-SU rate region with different values of RF-DC efficiency. }
\label{fig:rate:region}
\end{figure}

In Fig. \ref{fig:rate:region}, we  show the achievable rate region
 of the proposed  information and energy cooperation schemes and compare it with
 the conventional information cooperation only scheme \cite{Zheng_CCRN_13}.
We consider a scenario where the distances from the ST to all the
other terminals  are $1$m, while the distance from the PT to the
 PU is $2$m, therefore assistance from the ST is usually preferred by the
 PT. We assume that the ST has $4$ transmit antennas and all other terminals have a single antenna.   The primary   energy  is set to $20$ dB while the available
  secondary energy is $10$ dB. Path loss exponent is $3.5$ and the $K$ factor for the Rician channel model  is set to $5$ dB.
The RF-to-DC efficiency is equal to  $\eta=0.1, 0.5,$ and $1$.
It is seen that the achievable rate regions are greatly enlarged
thanks to the extra energy cooperation even with RF-to-DC
efficiencies as low as $\eta=0.1$. When the required PU rate is $2$ bps/Hz, the SU can
double or triple its rate compared to the case without energy
cooperation as $\eta$ varies from $0.1$ to $1$. When the SU rate is $1.5$
bps/Hz, the PU enjoys 75\% higher data rate when $\eta=1$.

The proposed additional energy cooperation clearly introduces a substantial
performance gain over the existing information cooperation only CR
scheme, and could be a promising solution for the future
CCRNs.

\section{Conclusion and future work}

This survey paper provides an overview of the SWIPT technology.  Different SWIPT techniques that split the received signal in orthogonal components have been discussed. We have shown that SWIPT introduces fundamental changes in the resource allocation problem and influences basic operations such as scheduling, power control, and interference management. Finally, a sophisticated CR network that enables information/energy cooperation between primary and secondary systems has been discussed as an example of new  SWIPT applications. SWIPT imposes many interesting and challenging new research problems and will be a key technology for the next-generation communication systems. In the following, we discuss some of the research challenges and potential solutions:
\begin{itemize}

\item Path loss:  The efficiency of SWIPT is expected to be unsatisfactory for long distance transmission unless  advanced resource allocation and antenna technology can be combined. Two possible approaches to overcome this problem include the use of massive MIMO and coordinate multipoint systems. The former increase the DoF offered to harvest energy and create highly directive energy/information beams steered towards the receivers. The later provides  spatial diversity  for combating  path loss by reducing the distance between transmitters and receivers. Besides, the distributed transmitters may be equipped with traditional energy harvesters (such as solar panels) and exchange their  harvested energy over a power grid so as to overcome potential energy harvesting imbalances in the network.

\item Communication and energy security: Transmitters can increase the energy of the
information carrying signal to facilitate energy harvesting at the receivers.  However, this may also increase their
susceptibility to eavesdropping due to the broadcast nature of wireless channels.  On the other hand, receivers requiring power transfer may take advantage of the transmitter by falsifying  their reported channel state information. Therefore, new QoS concerns on communication and energy security naturally arise in  SWIPT systems.

\item Hardware Development: Despite the wealth of theoretical techniques for SWIPT, so far, hardware implementations have mostly been limited to WPT systems that opportunistically harvest ambient energy. Thus, the development of SWIPT circuits is fundamental to investigate the tradeoff between SWIPT techniques, occurring due to inefficiencies of different circuit modules. For example, the TS technique is theoretically less efficient than PS, but the later suffers from power splitting losses that are not accounted for in theoretical studies.

\item Applications: SWIPT technology has promising applications in several areas that can benefit from ultra-low power sensing devices. Potential applications include structure monitoring by embedding sensors in buildings, bridges, roads, etc.,  healthcare  monitoring using implantable bio-medical sensors and building automation through smart sensors that monitor and control different building processes. However, for the successful realization of such SWIPT applications, several challenges have to be overcome at various layers from hardware implementation over protocol development to architectural design.
\end{itemize}

\section*{Acknowledgment}
This work was partially supported by the Research Promotion Foundation, Cyprus
under the project KOYLTOYRA/BP-NE/0613/04 ``Full-Duplex Radio: Modeling,
Analysis and Design (FD-RD)''.

\end{document}